\documentclass[a4paper,12pt]{article}
\usepackage{graphicx}
\usepackage{amsmath}
\usepackage{booktabs}
\usepackage{siunitx}
\usepackage{hyperref}
\usepackage{caption}
\usepackage{subcaption}
\usepackage{typearea}
\typearea{20}
\usepackage{multirow}

\providecommand{\keywords}[1]
{
  \small	
  \textbf{\textit{Keywords---}} #1
}
 
\title{Robustness of Proof of Team Sprint (PoTS) Against Attacks: A Simulation-Based Analysis}
\author{Naoki Yonezawa \thanks{\texttt{n.yonezawa@thu.ac.jp}. Faculty of Humanities and Social Sciences, Teikyo Heisei University, Japan.}}
\date{}

\begin{document}
\maketitle

\begin{abstract}
This study evaluates the robustness of Proof of Team Sprint (PoTS) against adversarial attacks through simulations, focusing on the attacker win rate and computational efficiency under varying team sizes (\( N \)) and attacker ratios (\( \alpha \)). Our results demonstrate that PoTS effectively reduces an attacker's ability to dominate the consensus process. For instance, when \( \alpha = 0.5 \), the attacker win rate decreases from 50.7\% at \( N = 1 \) to below 0.4\% at \( N = 8 \), effectively neutralizing adversarial influence. Similarly, at \( \alpha = 0.8 \), the attacker win rate drops from 80.47\% at \( N = 1 \) to only 2.79\% at \( N = 16 \). In addition to its strong security properties, PoTS maintains high computational efficiency. We introduce the concept of Normalized Computation Efficiency (NCE) to quantify this efficiency gain, showing that PoTS significantly improves resource utilization as team size increases. The results indicate that as \( N \) grows, PoTS not only enhances security but also achieves better computational efficiency due to the averaging effects of execution time variations. These findings highlight PoTS as a promising alternative to traditional consensus mechanisms, offering both robust security and efficient resource utilization. By leveraging team-based block generation and randomized participant reassignment, PoTS provides a scalable and resilient approach to decentralized consensus.
\end{abstract}


\section{Introduction}
Blockchain consensus mechanisms must be resilient against attacks while maintaining efficiency. Traditional mechanisms such as Proof of Work (PoW) and Proof of Stake (PoS) employ different strategies to secure the network. PoW enforces security through computational difficulty, requiring participants to solve cryptographic puzzles, thereby making attacks computationally infeasible. However, PoW is known for its high energy consumption and centralization tendencies due to mining pool dominance. PoS, in contrast, relies on economic incentives by granting block generation rights based on the stake held by participants, reducing energy consumption but introducing centralization risks and potential vulnerabilities such as the ``nothing-at-stake'' problem.

Proof of Team Sprint (PoTS) introduces an alternative approach to block generation by structuring consensus around randomly selected teams. Instead of individual nodes competing to solve a cryptographic challenge, PoTS assigns participants to teams, where each member generates a block in a coordinated manner. This design aims to distribute computational effort more efficiently while simultaneously increasing network security by requiring adversarial control over entire teams rather than individual participants. The randomized team assignment process further reduces the likelihood of an attacker consistently gaining control over the consensus process.

This paper investigates the robustness of PoTS against adversarial control by evaluating the probability of consecutive attacker wins and analyzing the computational efficiency of the protocol. By systematically varying team size \( N \) and attacker ratio \( \alpha \), we assess how PoTS mitigates the risk of prolonged adversarial influence. Through a simulation-based approach, we measure attacker success rates, total computation time, and the longest consecutive win streak of adversarial teams. These results provide insight into how PoTS enhances both security and efficiency compared to traditional consensus mechanisms.

The key contributions of this paper are as follows:
\begin{itemize}
    \item We develop a simulation framework to evaluate PoTS under varying attacker ratios and team sizes.
    \item We analyze the probability of attacker success in consecutive rounds, demonstrating how PoTS reduces adversarial influence as team size increases.
    \item We examine the impact of the synchronization mechanism in PoTS on computational efficiency and introduce the concept of Normalized Computation Efficiency (NCE) to quantify the relationship between security and performance.
    \item We compare simulation results with theoretical expectations, identifying discrepancies arising from PoTS's design and discussing their implications.
\end{itemize}

The remainder of this paper is structured as follows. Section~\ref{sec:related_work} reviews related work, summarizing key findings from existing studies on blockchain consensus mechanisms, attack models, and simulation-based security evaluations. Section~\ref{sec:simulation_model} details the simulation model, including network setup, computation time modeling, and attacker strategy. Section~\ref{sec:results} presents the simulation results, including attacker win rates, computation efficiency, and maximum consecutive attacker wins. Section~\ref{sec:discussion} discusses the implications of these results, highlighting the security advantages and efficiency characteristics of PoTS. Finally, Section~\ref{sec:conclusion} summarizes our findings and suggests directions for future research.

\section{Related Work and Background}
\label{sec:related_work}

To understand the robustness of PoTS, it is essential to compare it with existing consensus mechanisms and examine previous studies on blockchain security. Traditional blockchain consensus mechanisms, such as Proof of Work (PoW) and Proof of Stake (PoS), each have distinct strengths and vulnerabilities. PoTS introduces a new approach by leveraging team-based block generation, which impacts both security and computational efficiency.

\subsection{Comparison of PoW, PoS, and PoTS}

\begin{figure}[t]
\begin{center}
\includegraphics[scale=0.5]{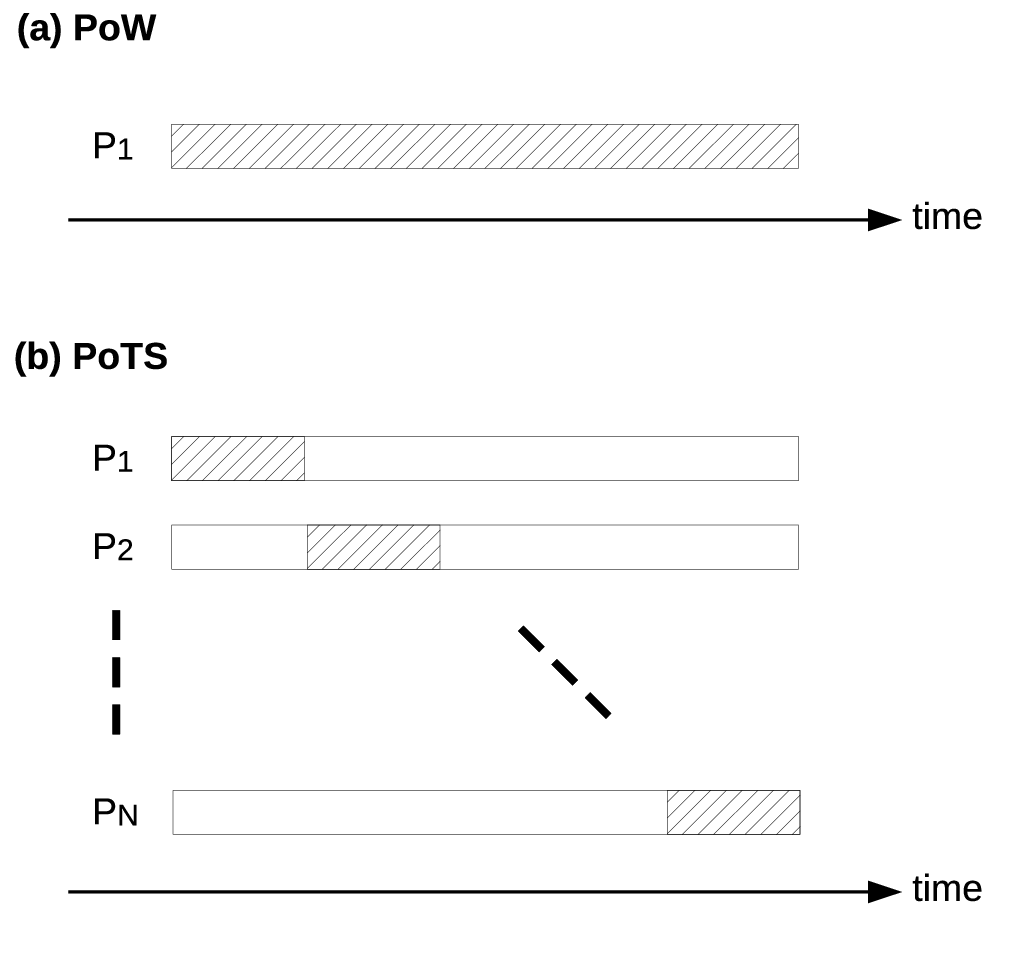}
\caption{Comparison of Block Generation Process in PoW and PoTS. Adapted from~\cite{yonezawa2024pots}.}
\label{fig:pow-pots-time-diagram}
\end{center}
\end{figure}

Blockchain consensus mechanisms are designed to ensure agreement among distributed participants while resisting adversarial control. PoW and PoS represent the two most widely adopted approaches, each with unique characteristics and associated risks.

PoW achieves consensus through computational difficulty, requiring participants to solve cryptographic puzzles to propose valid blocks. This mechanism ensures that block generation is probabilistic and favors nodes with higher computational power. While PoW is effective in maintaining decentralization, it is vulnerable to 51\% attacks, where an adversary controlling more than half of the total mining power can rewrite transaction history or prevent new transactions from being confirmed. Moreover, PoW suffers from high energy consumption and mining centralization due to the economic advantages held by large mining pools.

PoS, in contrast, determines block creation rights based on the amount of cryptocurrency staked by participants. Validators with a higher stake are more likely to be selected for block generation, reducing the energy-intensive computations required in PoW. However, PoS introduces risks of centralization, as wealthier participants accumulate more influence over time, reinforcing their control over the network. Additionally, PoS is susceptible to the Nothing-at-Stake problem, where validators can attempt to support multiple conflicting chains simultaneously without incurring significant penalties. This issue complicates fork resolution and raises concerns about security under adversarial conditions.

PoTS differs from both PoW and PoS by structuring consensus around teams rather than individual participants. Instead of a single node generating a block, PoTS assigns participants to randomly selected teams that work collaboratively to produce blocks in a sequential manner. As illustrated in Figure~\ref{fig:pow-pots-time-diagram}, this structure inherently increases security because an adversary must gain control of an entire team rather than a single node to manipulate the consensus process. As team size increases, the probability of an adversary successfully controlling a team decreases exponentially, making PoTS highly resistant to prolonged adversarial dominance. Additionally, the continuous reassignment of participants into new teams in each round prevents attackers from exploiting persistent group structures. 

Similar to PoW, PoTS includes a synchronization mechanism where teams complete block generation in a coordinated manner. This design enhances fairness and efficiency while maintaining a balance between security and resource utilization.

\subsection{Attack Resistance in Blockchain Consensus}

Security analyses of PoW and PoS have traditionally focused on several key attack vectors. One of the most studied threats is the probability of successful chain reorganizations, where an attacker attempts to override previously confirmed blocks by extending an alternative chain. PoW's security model relies on economic and computational constraints that make such attacks infeasible for adversaries with limited mining power. However, mining centralization can increase the likelihood of collusion-based attacks. In PoS, similar concerns arise when large stakeholders coordinate to override honest participants, leading to long-range attacks where old states of the blockchain can be rewritten.

Another significant consideration is Sybil attack resistance, where adversaries create a large number of fake identities to gain disproportionate influence. PoW mitigates Sybil attacks by tying participation to computational power, while PoS uses staking requirements as an economic barrier. PoTS inherently reduces the effectiveness of Sybil attacks by requiring adversaries to control entire teams rather than individual nodes, significantly increasing the cost and complexity of launching an attack.

Despite these security advantages, theoretical analyses of PoTS remain in their early stages. Unlike PoW and PoS, where formal security models have been extensively studied, PoTS's novel team-based structure requires further validation. The probabilistic nature of team formation and the sequential dependency of block generation introduce unique security considerations that differ from traditional consensus mechanisms. Given these complexities, simulation-based validation is necessary to quantify PoTS's resistance to adversarial strategies and assess its real-world applicability. This study contributes to this effort by systematically evaluating the probability of attacker success and demonstrating how PoTS enhances both security and computational efficiency.

\subsection{Related Work}

Simulation-based studies have been widely used to evaluate various blockchain consensus mechanisms, particularly Proof-of-Work (PoW) and Proof-of-Stake (PoS), focusing on security, performance, and economic feasibility. In this section, we summarize key findings from existing studies relevant to the evaluation of PoTS.

One of the primary areas of blockchain simulation research is attack success probability analysis under different computational power distributions. Many studies have investigated the impact of adversarial strategies such as the 51\% attack \cite{sayeed2019assessing, amin202051, bastiaan2015preventing}, selfish mining \cite{grunspan2018profitability, leelavimolsilp2018preliminary, feng2019selfish}, and Sybil attacks \cite{platt2021sybil, rajabi2023feasibility}. These works provide valuable insights into how computational power is distributed among participants and how malicious actors may exploit network vulnerabilities to gain unfair advantages. For instance, research on Sybil attacks in identity-augmented PoS networks suggests that malicious entities can manipulate identity-based selection mechanisms to increase their chances of forging blocks \cite{platt2021sybil}. Similarly, Sybil attack feasibility in shard-based permissionless blockchains has been examined using probabilistic models to determine the conditions under which attackers can gain control over specific shards \cite{rajabi2023feasibility}. These findings are particularly relevant to PoTS, where Sybil attacks could influence the integrity of team formation.

Another important aspect of blockchain security research involves analyzing the effects of block generation difficulty adjustments. Works such as \cite{gervais2016security, rebello2022security, yang2019effective} have studied how difficulty retargeting mechanisms influence the stability and fairness of PoW and PoS networks. In particular, an effective scheme against 51\% attacks based on history-weighted difficulty adjustments was proposed in \cite{yang2019effective}, demonstrating how historical mining participation can be leveraged to deter adversarial control. Additionally, studies on selfish mining in Ethereum \cite{feng2019selfish} highlighted the impact of uncle block rewards on mining strategies, revealing that Ethereum exhibits a lower threshold for profitable selfish mining compared to Bitcoin. Recent work on stubborn mining and its combination with Eclipse attacks \cite{nayak2016stubborn} further generalizes selfish mining strategies and provides insights into the potential threats PoTS must mitigate.

Beyond security evaluations, blockchain simulations have also been employed to assess performance and fairness across various consensus mechanisms. Comparative analyses of PoW, PoS, Delegated PoS (DPoS), and Practical Byzantine Fault Tolerance (PBFT) protocols have been conducted in \cite{kaur2021mbcp, fan2020performance}, providing insights into the trade-offs between decentralization, energy efficiency, and transaction throughput. The importance of fairness in blockchain consensus, especially within permissioned blockchain networks, has been analyzed in \cite{malakhov2021use}, where a sliding-window PoW mechanism was introduced to ensure a more equitable distribution of mining rewards among participants. This concept of fairness enhancement is relevant to PoTS, where ``team-based sequential and restricted consecutive mining'' serves a similar purpose.

Simulation methods have also been instrumental in incorporating economic models into blockchain analysis. Research on the risk preferences of miners and their implications for network security \cite{civitarese2018risk} suggests that risk-averse miners may reduce network security by leaving during periods of high volatility, whereas risk-seeking miners contribute to long-term stability. Additionally, market-driven studies such as \cite{shanaev2019cryptocurrency} have analyzed how 51\% attacks impact cryptocurrency value, demonstrating significant price fluctuations following such attacks.

The reviewed studies highlight a variety of simulation techniques used for blockchain evaluation. Markov chain models have been employed to assess attack success probabilities and long-term network stability \cite{bastiaan2015preventing, rebello2022security}, while agent-based simulations have been widely used for security and performance evaluations \cite{serena2022security}. Blockchain simulation frameworks such as BlockSim \cite{alharby2019blocksim} and PoW network simulators \cite{wuthier2021proof} have facilitated the reproducibility of blockchain behavior under controlled experimental conditions. Recent advancements in blockchain network simulators \cite{wuthier2021proof} provide valuable tools for constructing a PoTS-specific simulation environment.

While these studies provide a strong foundation for evaluating PoW and PoS consensus mechanisms, they do not fully address the unique challenges of modeling team-based consensus approaches such as PoTS. The PoTS framework introduces new dimensions to blockchain simulation, including team formations, inter-team competition, and cooperative block validation. Existing studies on cooperative mining strategies, such as Robust PoS (RPoS) \cite{li2020robust}, provide some preliminary insights into coordinated mining behavior. Additionally, competitive selfish mining models \cite{azimy2019competitive} offer a game-theoretic perspective on multi-group competition, which is a crucial aspect of PoTS. Further research is needed to model adversarial strategies in PoTS environments and evaluate their long-term security implications.

In summary, existing blockchain simulation studies have laid essential groundwork for understanding consensus security, performance, and economic feasibility. However, PoTS-specific challenges such as modeling team-based strategies and evaluating adversarial behavior over multiple rounds remain open areas of research. Future work should integrate insights from traditional blockchain simulations while developing new methodologies tailored to the unique characteristics of PoTS.

\section{Simulation Model}
\label{sec:simulation_model}

To evaluate the security and efficiency of PoTS, we developed a simulation model that systematically varies the attacker ratio \( \alpha \) and team size \( N \). The model simulates multiple rounds of consensus, where participants are dynamically assigned to teams to generate blocks. The simulation tracks various performance metrics, including attacker win rate, total computation time, and the longest streak of consecutive attacker wins.

\subsection{Simulation Parameters}
The simulation is configured with the following parameters:
\begin{itemize}
    \item \textbf{Total nodes:} The network consists of \( 1,600 \) nodes participating in the PoTS consensus process.
    \item \textbf{Rounds:} Each simulation runs for \( 1,000 \) consensus rounds.
    \item \textbf{Team workload time:} The total workload assigned to a team is set to a fixed value of \( 600 \) seconds. Consequently, the base computation time per node is given by:
    \[
    T_{\text{base}} = \frac{T_{\text{workload}}}{N} = \frac{600}{N}.
    \]
    \item \textbf{Attacker ratio \( \alpha \):} The fraction of nodes controlled by an attacker varies from \( 0.0 \) to \( 1.0 \) in increments of \( 0.1 \) (i.e., \( 11 \) different values).
    \item \textbf{Team sizes \( N \):} We evaluate PoTS for \( N = 1, 2, 4, 8, 16, 32, 64 \) (i.e., \( 7 \) different values). Notably, when \( N = 1 \), PoTS operates equivalently to PoW.
    \item \textbf{Repetitions:} Each simulation configuration (i.e., a fixed \( N \) and \( \alpha \) pair) is repeated \( 100 \) times to ensure statistical robustness.
\end{itemize}

\subsection{Network Setup}
\label{subsec:network_setup}
As mentioned earlier, the simulated network consists of \( n = 1,600 \) participants, each representing an individual node in the PoTS consensus process. At the beginning of each round, participants are randomly grouped into teams of fixed size \( N \). The number of teams per round is given by:

\[
M = \frac{n}{N}
\]

assuming \( n \) is an integer multiple of \( N \). These teams are reassembled randomly at the start of each new round to prevent adversarial nodes from consistently collaborating.

A fraction \( \alpha \) of the total computational power is controlled by an adversary. These attacker-controlled nodes are distributed randomly across teams in the same manner as honest nodes, ensuring that adversarial influence is spread across the network rather than concentrated in a specific subset of teams.

\subsection{Computation Time Model}
\label{subsec:computation_time_model}
Each node generates a block independently, with its execution time determined by a computation time model. The block generation time \( T_i \) for a participant \( i \) is given by:

\[
T_i = T_{\text{base}} \cdot \gamma_i
\]

where \( T_{\text{base}} \) is the base computation time per node, and \( \gamma_i \) is a random multiplier sampled from a uniform distribution within the range \((0.8, 1.2)\). This stochastic variation introduces fluctuations in execution time, modeling real-world variations in computational performance.

Within each team, nodes generate blocks sequentially. The total execution time of a team is determined by the sum of the individual block generation times of its members. The team that completes its block sequence first determines the round's global completion time \( T_{\text{round}} \), given by:

\[
T_{\text{round}} = \min_{\text{all teams}} \sum_{j=1}^{N} T_j
\]

where \( T_j \) represents the execution time of the \( j \)-th node within a team. Once the fastest team completes its block sequence, the round ends, and all other teams must also terminate at the same time. This synchronization ensures that the next consensus round begins uniformly across the network.

To implement this mechanism, we apply the following process:
\begin{enumerate}
\item Each node within a team records its execution time for generating a block.
\item The fastest team determines \( T_{\text{round}} \).
\item Any team that has not completed its work by \( T_{\text{round}} \) is required to stop.
\item For nodes in these teams:
\begin{itemize}
\item If their cumulative execution time exceeds \( T_{\text{round}} \), they are required to terminate their computations at \( T_{\text{round}} \).
\item Nodes that have not started their assigned computation are assigned zero execution time.
\end{itemize}
\end{enumerate}

This synchronized execution mechanism in PoTS ensures fairness and consistency in block generation. By aligning execution times across teams, it makes the recorded computation time more reflective of real-world conditions. As a result, the theoretical efficiency gains expected from increasing \( N \) are adjusted accordingly, as seen in the Normalized Computation Efficiency (NCE) results, where the observed efficiency follows a more realistic trend.

\subsection{Attacker Strategy}

The attacker aims to maximize consecutive successful block generations by controlling an entire team within a round. The probability of an attacker fully controlling a team is given by:

\[
P_{\text{attack}}(N, \alpha) = \alpha^N
\]

where \( \alpha^N \) represents the probability that all \( N \) members of a randomly formed team are controlled by the attacker. Since teams are reformed in each round, the attacker must repeatedly achieve full control of a team to sustain a long winning streak.

The longest sequence of consecutive attacker wins is a key metric for evaluating the security of PoTS. We aim to analyze how increasing the team size \( N \) reduces the probability of an attacker maintaining a long winning streak. As team size grows, the likelihood of an attacker fully controlling a team is expected to decrease significantly, making sustained dominance less feasible.

This model provides a quantitative framework for assessing PoTS's resilience against adversarial control. By examining attacker win rates and consecutive win streaks under various values of \( N \) and \( \alpha \), we evaluate how PoTS mitigates the risk of prolonged adversarial influence while maintaining computational efficiency.

\section{Simulation Results}
\label{sec:results}

This section presents the results of our simulation, which evaluates the robustness of Proof of Team Sprint (PoTS) under different attacker ratios and team sizes. Three key aspects are analyzed: the attacker win rate, the computation efficiency, and the longest streak of consecutive attacker wins.

\subsection{Comparison of Simulation and Theoretical Predictions}

\subsection{Comparison of Simulation and Theoretical Predictions}

To assess the accuracy of our simulation model, we compare the observed attacker win rates with theoretical predictions for different team sizes \(N\) and attacker ratios \(\alpha\). Tables~\ref{tab:sim_vs_theory_1} and \ref{tab:sim_vs_theory_2} summarize the results for three representative values of \(\alpha\): 0.2, 0.5, and 0.8. Table~\ref{tab:sim_vs_theory_1} presents the simulation results for \(\alpha = 0.2\) and \(\alpha = 0.5\), while Table~\ref{tab:sim_vs_theory_2} provides the corresponding results for \(\alpha = 0.8\). Each table includes the mean and standard deviation of the simulated attacker win rates, computed from 100 independent simulation runs.

The results in Tables~\ref{tab:sim_vs_theory_1} and \ref{tab:sim_vs_theory_2} show that for small team sizes (\(N = 1, 2, 4\)), the simulated attacker win rates closely match the theoretical expectations but exhibit relatively high variance. As \(N\) increases, the standard deviation of the simulation results decreases, and the mean values become more closely aligned with theoretical predictions. This trend suggests that larger team sizes lead to more stable outcomes, reducing the impact of stochastic variations in computation times. The implications of this trend are further discussed in Section~\ref{sec:discussion}.

\begin{table}[h]
    \centering
    \caption{Comparison of simulated and theoretical attacker win rates for \(\alpha = 0.2\) and \(\alpha = 0.5\) across different team sizes.}
    \label{tab:sim_vs_theory_1}
    \begin{tabular}{c|ccc|ccc}
        \toprule
        \multirow{2}{*}{Team Size \(N\)} & \multicolumn{3}{c|}{\(\alpha = 0.2\)} & \multicolumn{3}{c}{\(\alpha = 0.5\)} \\
        & Sim. Mean & Sim. Std & Theory & Sim. Mean & Sim. Std & Theory \\
        \midrule
        1  & 20.32 & 5.87 & 20.0  & 50.70 & 5.89 & 50.0  \\
        2  & 3.87  & 1.04 & 4.0   & 24.99 & 2.83 & 25.0  \\
        4  & 0.17  & 0.13 & 0.16  & 6.22  & 1.04 & 6.25  \\
        8  & 0.001 & 0.01 & 0.0003 & 0.40  & 0.21 & 0.39  \\
        16 & 0.0   & 0.0  & $10^{-9}$  & 0.0   & 0.0  & $1.5 \times 10^{-3}$  \\
        32 & 0.0   & 0.0  & $10^{-21}$ & 0.0   & 0.0  & $2.3 \times 10^{-8}$  \\
        64 & 0.0   & 0.0  & $10^{-43}$ & 0.0   & 0.0  & $5.4 \times 10^{-18}$ \\
        \bottomrule
    \end{tabular}
\end{table}

\begin{table}[h]
    \centering
    \caption{Comparison of simulated and theoretical attacker win rates for \(\alpha = 0.8\) across different team sizes.}
    \label{tab:sim_vs_theory_2}
    \begin{tabular}{c|ccc}
        \toprule
        \multirow{2}{*}{Team Size \(N\)} & \multicolumn{3}{c}{\(\alpha = 0.8\)} \\        
         & Sim. Mean & Sim. Std & Theory \\
        \midrule
        1  & 80.47 & 5.34 & 80.0 \\
        2  & 64.06 & 3.88 & 64.0 \\
        4  & 40.93 & 3.39 & 40.96 \\
        8  & 17.04 & 1.70 & 16.78 \\
        16 & 2.79  & 0.61 & 2.81  \\
        32 & 0.07  & 0.08 & 0.079  \\
        64 & 0.0   & 0.0  & $6.3 \times 10^{-5}$ \\
        \bottomrule
    \end{tabular}
\end{table}

\subsection{Attacker Win Rate}

\begin{figure}[h]
    \centering
    \includegraphics[width=0.8\linewidth]{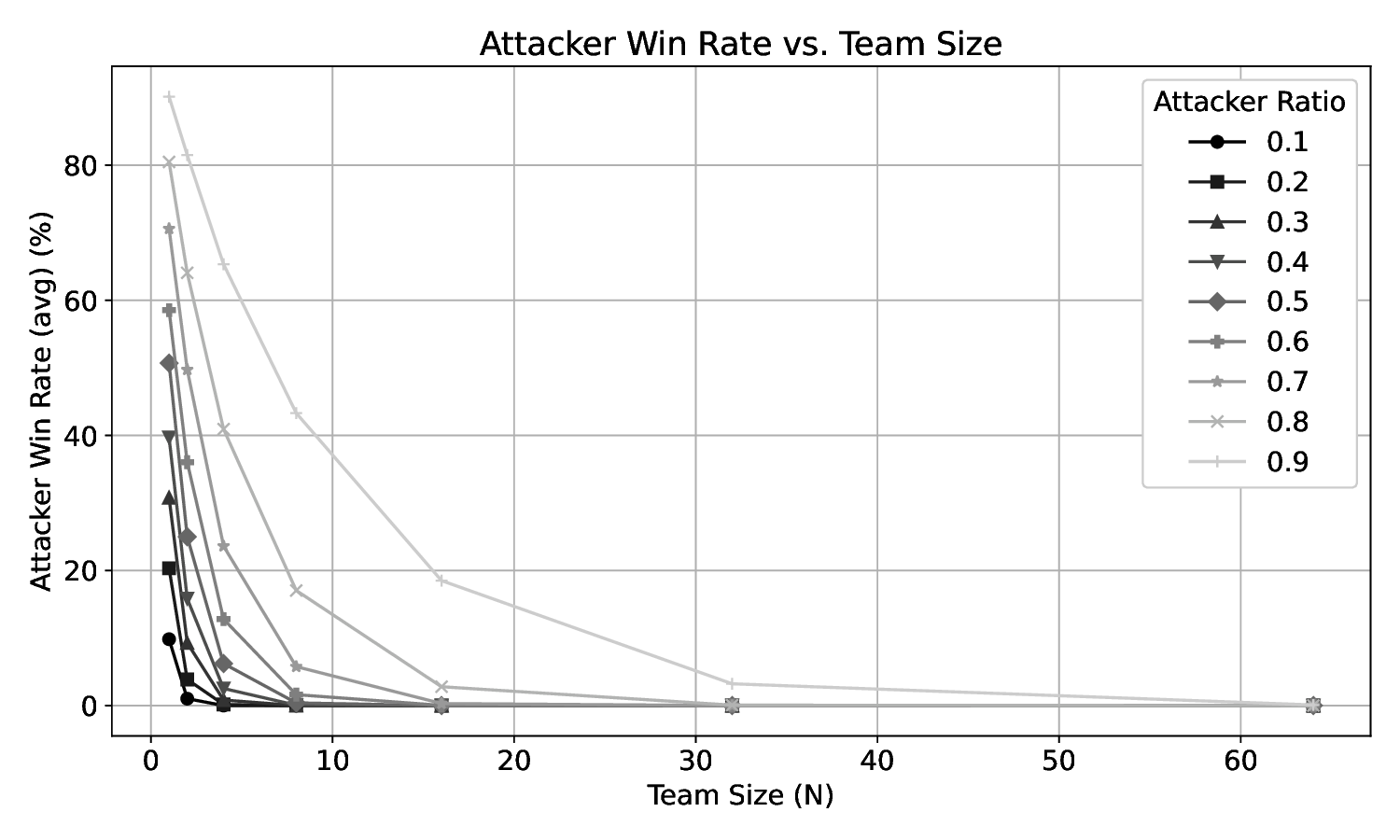}
    \caption{Attacker win rate vs. team size for different attacker ratios. Larger teams significantly reduce the attacker's success probability.}
    \label{fig:attacker_win_rate}
\end{figure}

The attacker win rate is a crucial metric that represents the probability of an adversary successfully gaining control over the consensus process. Figure~\ref{fig:attacker_win_rate} illustrates how the attacker win rate varies as a function of team size for different attacker ratios. Note that results for \(\alpha = 0.0\) and \(\alpha = 1.0\) are omitted from the figure, as they are trivial cases: when \(\alpha = 0.0\), the attacker always loses, and when \(\alpha = 1.0\), the attacker always wins.

The results demonstrate that when \(N\) is small, the probability of attacker success aligns well with theoretical predictions. However, as \(N\) increases, the attacker's ability to dominate the consensus process diminishes rapidly. For \(\alpha \leq 0.5\), the win rate becomes negligible for \(N \geq 16\), effectively neutralizing adversarial influence. When \(\alpha \geq 0.8\), the attacker still maintains a moderate probability of success at small \(N\), but this probability declines significantly as \(N\) grows.

\subsection{Normalized Computation Efficiency (NCE)}

To evaluate the computational efficiency of PoTS, we introduce the concept of Normalized Computation Efficiency (NCE), which is defined as the reciprocal of the total computation time:

\[
\text{NCE}(N) = \frac{T_1}{T_N} 
\]

where \( T_N \) represents the total computation time for a given team size \(N\). A higher NCE value indicates improved energy efficiency, as it implies that less time is required to complete consensus.

\begin{figure}[h]
    \centering
    \includegraphics[width=0.8\linewidth]{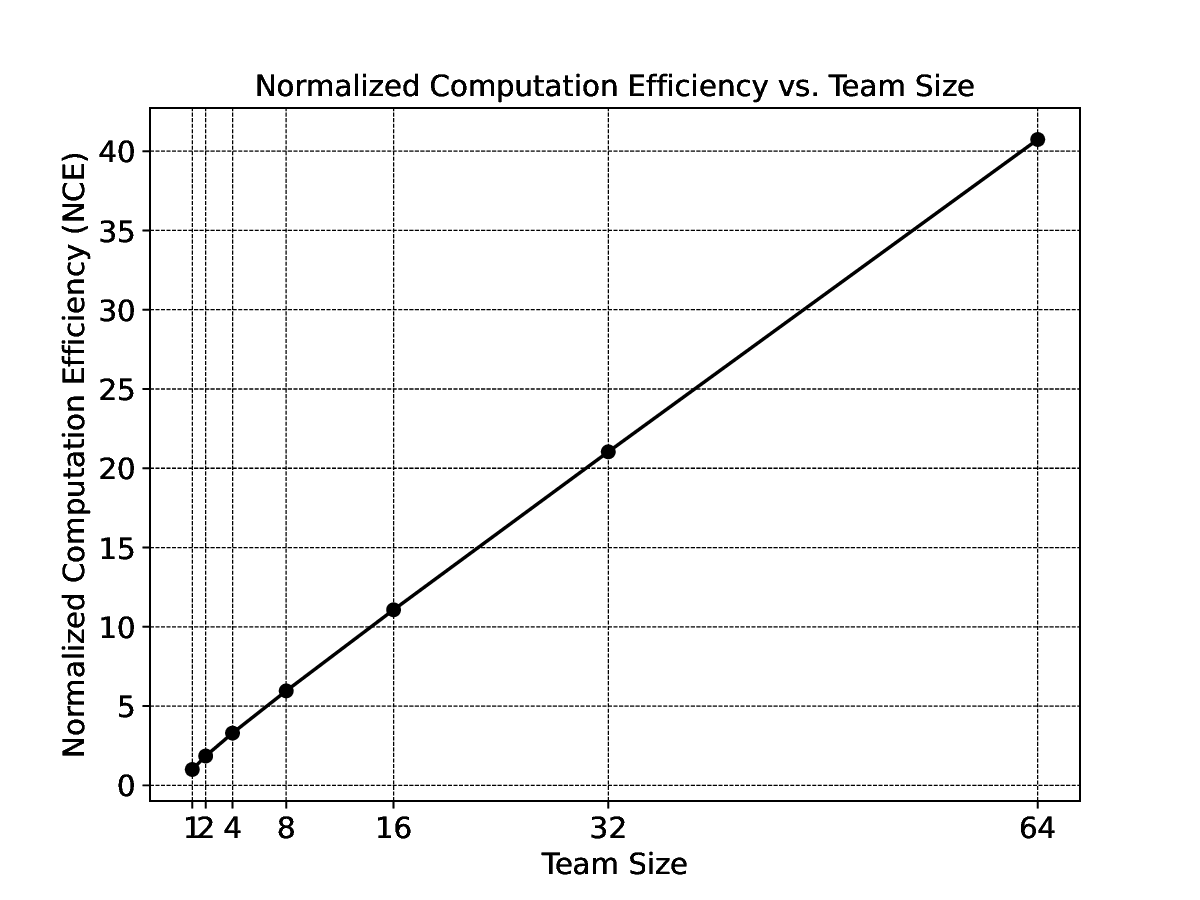}
    \caption{Normalized Computation Efficiency (NCE) vs. team size for attacker ratio \( \alpha = 0.5 \). Higher values indicate better energy efficiency.}
    \label{fig:nce}
\end{figure}

Figure~\ref{fig:nce} presents the NCE values measured for different team sizes under an attacker ratio of \( \alpha = 0.5 \). We selected \( \alpha = 0.5 \) as the representative case because the identity of the fastest team is independent of the attacker ratio. Since block generation is determined primarily by individual node execution times rather than attacker presence, the overall trend observed at \( \alpha = 0.5 \) is expected to generalize to other values of \( \alpha \).

The results confirm that NCE increases with \(N\), indicating improved efficiency. However, the efficiency gain does not scale perfectly with \(N\). This deviation arises from the stochastic variation factor \( \gamma_i \) introduced in Section~\ref{subsec:network_setup}, which affects execution times at the node level. When \(N\) is small, these variations have a stronger global impact, making the total execution time at \( N = 1 \) significantly shorter than \( 64 \) times that at \( N = 64 \). As \( N \) increases, the influence of \( \gamma_i \) diminishes due to averaging effects, leading to more predictable computation times.

\subsection{Maximum Consecutive Attacker Wins}
\label{subsec:maximum_consecutive_attacker_wins}

\begin{figure}[h]
    \centering
    \includegraphics[width=0.8\linewidth]{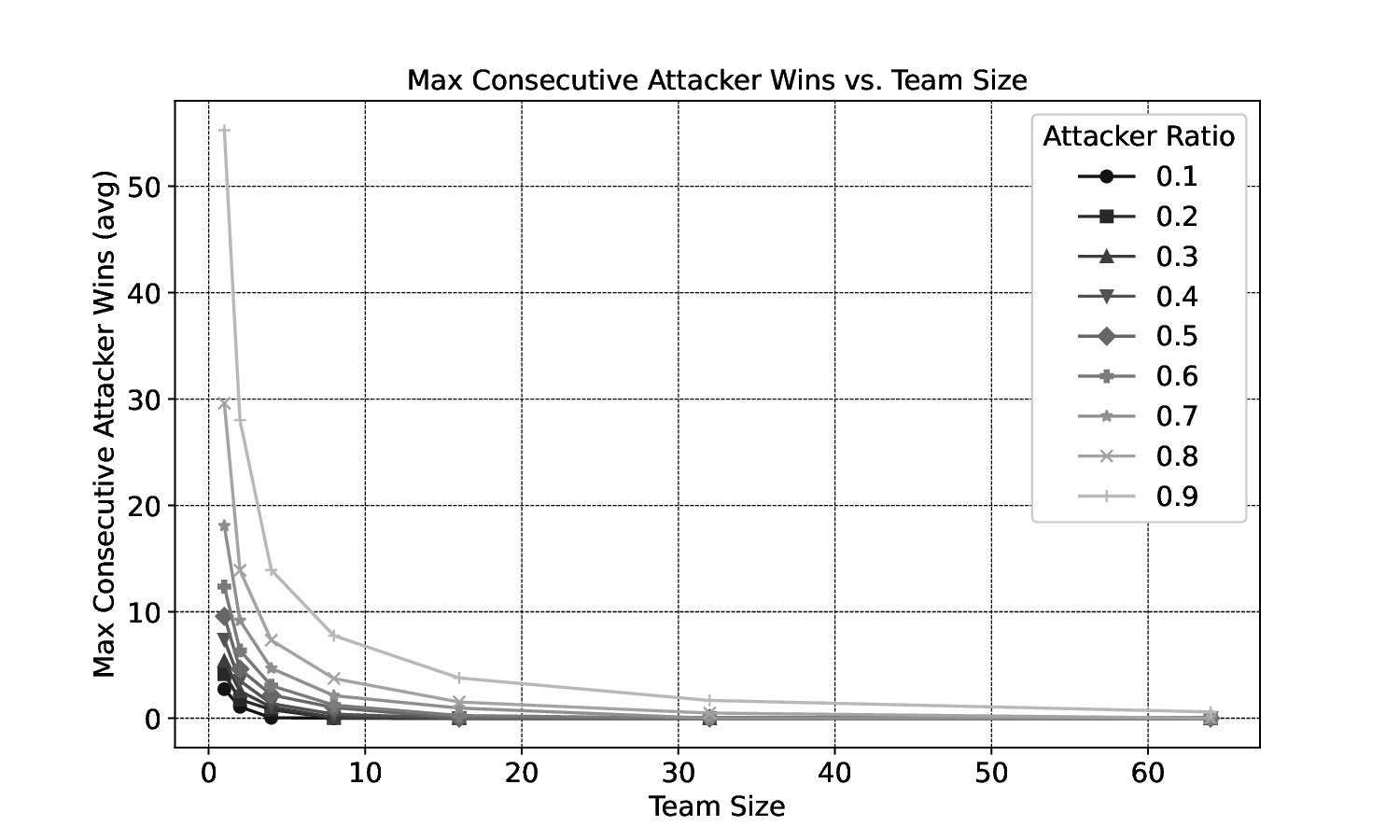}
    \caption{Maximum consecutive attacker wins vs. team size for different attacker ratios. Larger teams significantly limit an attacker's ability to sustain consecutive wins.}
    \label{fig:max_attacker_streak}
\end{figure}

Figure~\ref{fig:max_attacker_streak} presents the maximum consecutive attacker wins observed in the simulations. The results indicate that for small \(N\), an attacker can sustain prolonged winning streaks, particularly when \(\alpha\) is high. However, as \(N\) increases, the maximum streak length drops sharply, reinforcing the robustness of PoTS. 

\section{Discussion}
\label{sec:discussion}

The results presented in the previous section provide valuable insights into the behavior of Proof of Team Sprint (PoTS) under various attacker ratios and team sizes. In this section, we discuss the trends observed in the simulation results compared to theoretical expectations, and we analyze the broader security implications of our findings.

\subsection{Comparison Between Simulation and Theory}

While the theoretical model provides a useful baseline for predicting attacker win rates and computational efficiency, the simulation results confirm that PoTS follows theoretical predictions closely, particularly as team size increases. The alignment between simulation and theory becomes more evident for larger team sizes, where the impact of individual node execution time variability is mitigated by averaging effects.

One factor influencing the observed results is the random multiplier applied to computation times. In the simulation, each node's execution time is scaled by a random factor within the range \((0.8, 1.2)\), introducing variability in block generation times. This randomness affects both the observed attacker win rates and the normalized computation efficiency (NCE). However, as team size increases, the averaging effect reduces the influence of this variability, leading to more stable and predictable outcomes.

Another key factor is the impact of individual execution time variability on efficiency scaling. As discussed in Section~\ref{subsec:computation_time_model}, when \( N \) is small, the total execution time is more sensitive to the stochastic variation in individual node speeds. This effect prevents execution time at \( N = 1 \) from being exactly \( 64 \) times longer than that at \( N = 64 \), as variations in \( \gamma_i \) influence global timing more strongly when fewer nodes are involved. As \( N \) increases, this variability is averaged out, making the total execution time more predictable. This explains why the NCE does not scale perfectly linearly with \( N \), but still shows clear improvements as team size grows.

Furthermore, the results in Tables~\ref{tab:sim_vs_theory_1} and \ref{tab:sim_vs_theory_2} indicate that for small team sizes, the simulated attacker win rates closely align with theoretical expectations. As \(N\) increases, the probability of attacker success in the simulation remains consistent with theory, demonstrating that PoTS effectively neutralizes adversarial influence when \(N\) is sufficiently large. Additionally, the randomized team selection process appears to contribute to increased security by limiting an attacker's ability to maintain sustained control over the consensus process across multiple rounds.

\subsection{Security Implications}

The simulation results confirm that PoTS provides significant security advantages over traditional consensus mechanisms. One of the most critical observations is that PoTS effectively limits an attacker's ability to sustain long winning streaks, thereby reducing the risk of prolonged adversarial dominance.

The maximum consecutive attacker wins analysis reinforces this finding. When the team size is small, an attacker with a high computational fraction (\(\alpha \geq 0.8\)) can achieve extended winning streaks, posing a potential security threat. However, as \(N\) increases, the length of the longest attacker win streak rapidly declines. This demonstrates that PoTS naturally mitigates the risk of blockchain takeover, ensuring that adversarial control cannot persist indefinitely. Even when the attacker ratio is as high as \(\alpha = 0.9\), the results show that an attacker cannot consistently maintain control once team sizes exceed a certain threshold.

Another key security advantage of PoTS is that increasing the team size reduces the probability of full-team control by an attacker. Since block generation in PoTS requires sequential contributions from all members of a team, an attacker must control an entire team in order to successfully override the honest majority. As demonstrated in the attacker win rate results, for moderate values of \(N\), such as \(N \geq 16\), the probability of attacker success becomes negligible, even for relatively high values of \(\alpha\). This suggests that PoTS provides strong resistance against 51\% attacks, as an attacker must control a significantly greater portion of the network than in traditional PoW-based systems.

The randomized team selection process further strengthens PoTS's security by preventing an attacker from consistently exploiting the same set of compromised nodes. Unlike PoW, where mining power is continuously applied to the same chain, PoTS dynamically redistributes participants across teams in each round. As a result, even if an attacker successfully controls a team in one round, they are unlikely to replicate this control in subsequent rounds. This significantly reduces the feasibility of long-term strategic attacks.

Finally, our findings suggest that PoTS effectively enhances both security and computational efficiency. While increasing the team size strengthens security by reducing the probability of an attacker gaining control, it also influences execution dynamics due to the stochastic variation in individual node speeds. However, as evidenced by the NCE results, PoTS maintains high computational efficiency while significantly improving security. These findings highlight the importance of selecting an optimal team size to maximize both security and performance in real-world implementations of PoTS-based blockchain systems.

\section{Conclusion}
\label{sec:conclusion}

This study investigated the security and computational efficiency of Proof of Team Sprint (PoTS) as a consensus mechanism. Through a simulation-based approach, we evaluated how PoTS mitigates adversarial influence by distributing computational work across randomly formed teams. By systematically varying team size \( N \) and attacker ratio \( \alpha \), we quantified PoTS's resilience against prolonged adversarial control and analyzed its computational efficiency.

The results demonstrate that PoTS effectively limits the probability of consecutive attacker wins, particularly as team size increases. Unlike traditional PoW-based systems, where an attacker's success is directly proportional to their computational power, PoTS introduces an additional layer of security by requiring full-team control. The probability of adversarial dominance decreases exponentially with increasing \( N \), making large-team configurations significantly more resistant to sustained attacks. Furthermore, the randomized reassignment of team members in each round prevents an attacker from exploiting persistent group structures, further enhancing network security.

In addition to security benefits, PoTS maintains high computational efficiency while reducing unnecessary energy consumption. The NCE analysis confirms that PoTS scales efficiently with increasing \( N \), although the efficiency gains do not follow a perfect linear trend. This deviation arises from the stochastic variation in individual node execution times, which has a greater influence at smaller \( N \). As \( N \) increases, the averaging effect reduces this impact, making PoTS both scalable and computationally efficient.

This research provides a foundation for further exploration of PoTS-based consensus mechanisms. Several key areas for future work remain:
\begin{itemize}
    \item \textbf{Refining the simulation model}: While our simulations capture essential dynamics, incorporating more realistic computational constraints and communication delays could enhance accuracy.
    \item \textbf{Exploring dynamic network conditions}: Our current model assumes fixed team sizes and attacker ratios. Future studies should investigate scenarios with fluctuating participation rates, adaptive attacker strategies, and changing network topologies.
    \item \textbf{Increasing the number of simulation rounds}: Extending the simulation duration would provide deeper insights into long-term trends, particularly in adversarial settings with evolving attack strategies.
    \item \textbf{Comparing PoTS with alternative consensus mechanisms}: While our study includes an implicit comparison with PoW through the \( N = 1 \) case, further research is needed to evaluate PoTS against PoS and other emerging consensus models. Analyzing its relative security, efficiency, and scalability in diverse blockchain architectures would provide deeper insights into its practical applications and advantages.    \item \textbf{Investigating real-world implementation strategies}: Future research should explore practical deployment considerations, such as network latency effects, real-time team formation mechanisms, and incentive structures for PoTS-based blockchains.
\end{itemize}

By improving our understanding of PoTS's strengths and limitations, future research can help optimize its design for real-world applications, ensuring both robust security and sustainable efficiency in decentralized systems.

\section*{Acknowledgments}
The author declares no specific funding for this research and wishes to acknowledge his institution for providing a supportive research environment.

\bibliographystyle{IEEEtran}
\bibliography{pots_robust}

\end{document}